\documentclass[10pt,conference]{IEEEtran}
\usepackage{cite}
\usepackage{amsmath,amssymb,amsfonts}
\usepackage{algorithmic}
\usepackage{graphicx}
\usepackage{textcomp}
\usepackage{xcolor}
\def\BibTeX{{\rm B\kern-.05em{\sc i\kern-.025em b}\kern-.08em
    T\kern-.1667em\lower.7ex\hbox{E}\kern-.125emX}}
\usepackage{enumitem} % for unidented numbered lists 
\usepackage{xspace}
\newcommand{\xn}{\noindent}
\newcommand{\xv}{\vspace{0.1cm}}

\newcommand{\au}{autonomous\xspace}
\newcommand{\depsys}{depicted system\xspace}

\newcommand{\aisk}{AI Skill\xspace}
\begin{document}

\title{Toward Methods and Tools for Critical Examination of Disparate Systems for Discovery of Hidden Dependencies in Engineering of Autonomous and Semi-Autonomous Systems}

\author{\IEEEauthorblockN{Assaf Marron}
\IEEEauthorblockA{\textit{Department of Computer Science and Applied Mathematics} \\
\textit{Weizmann Institute of Science}\\
Rehovot, Israel \\
assaf.marron@weizmann.ac.il}
\and
\IEEEauthorblockN{David Harel}
\IEEEauthorblockA{\textit{Department of Computer Science and Applied Mathematics} \\
\textit{Weizmann Institute of Science}\\
Rehovot, Israel \\
david.harel@weizmann.ac.il}
}

\maketitle

\begin{abstract}
We present a methodology, an initial AI Skill package, and emerging results, for enhancing software engineers' acquisition and application of domain knowledge and system-wide engineering perspectives towards the development of an \au or semi-\au system in a domain that may be new to them.
At the core of the approach is a process involving proactive adversarial critiquing of videos of disparate systems in related environments. Among other things, the engineers and automated AI tools look for interactions, dependencies, and environment assumptions manifested in the observed systems, which are applicable also to the target system, but that otherwise might remain unarticulated in project artifacts. The approach and its supporting tools can enhance the quality of the target system, the individual engineer's skills and the organization's engineering practices.
\end{abstract}


%%%%%%%%%%%%%%%%%%%%%%%%%%%%%%%%%%%%%%%%
\section{Introduction and Overview}\label{sec:intro}
%%%%%%%%%%%%%%%%%%%%%%%%%%%%%%%%%%%%%%%%

Development of \au and semi-\au systems,
such as delivery vehicles and drones, 
autonomous agricultural machines,
industrial robotic arms,
and floor-cleaning robots, 
requires software engineers to apply system-wide considerations, covering software, hardware, behavior of users and other humans, related systems, the physical environment, etc., even when carrying out seemingly encapsulated tasks ~\cite{leveson1986softwareEngSysView,aniculaesei2018holisticSE,goues2024softwareEngSystemView}.
The implication of required learning, training and orientation has been addressed mainly in discussions of requirements engineering (RE)~\cite{araujo2025domainKnowledgeRE,beyer1995apprenticingRE,fischer1998evolvingKnowledge}. 

Complementing RE-driven learning, acquiring a broad understanding of the domain and related environments can be supported by the vast amounts of available information, internal and public: systems specifications, user documentation, training materials, engineering and domain-specific textbooks, software code, procurement RFPs, published domain lore about challenges in the development and use of various systems, and more. Furthermore, 
such information can be digested and organized using modern AI tools. 

Here we propose extending such classical studying and acquiring of domain knowledge, with a more exploratory and inquisitive learning approach. Beyond adding a proactivity dimension, this methodology focuses on discovering interactions and interdependencies between systems and between systems and their environment that may otherwise remain unarticulated when specifying a target system, testing and validating it, or planning its deployment in diverse environments.  

The methodology calls for engineer-driven AI-assisted critical examination of videos\footnote{Regarding the term \emph{video}: We could not find a more appropriate term that signifies sequences of still images captured by a camera over a continuous interval of time, depicting a scene or a scenario, optionally accompanied by audio; we chose \emph{video} over terms like movie, clip, movie clip, etc.}
of disparate systems, each of which may offer knowledge and considerations relevant to the development project at hand. The engineers and AI tools are directed to proactively look in these videos for emergent phenomena, absence of elements, actualized or potential interactions with other systems and environment elements, etc.  

The videos to be inspected may be obtained from diverse sources, including, e.g., ones taken by the engineer when observing real-world systems, prototypes or simulations; publicly available videos, posted for personal, commercial, or educational purposes; and videos from supervision or security cameras. 
A given video may focus on the system that one is interested in, or it may show it as part of the environment of other entities or events. 

We focus on videos of systems, as they complement other materials
\footnote{For discussion of search for hidden assumptions in specifications and models see~\cite{MarronAssmannLimonad2025EtAlHiddenAssumptions}.}
 in several ways:
First, videos can contain extensive amounts of information, especially about interactions; consider an agricultural device interacting with plants, soil, air, other equipment and humans, or an \au in-hospital delivery cart interacting with humans in the hallway by avoiding them or by eliciting reactions.
Second, compared with training through hands-on experimentation, apprenticeship or watching live demonstrations, a video may be replayed, enabling close examination of details that include co-occurring interactions that could not be fully absorbed in real time.
Third, in contrast to code or document inspections, we believe that observers of a video can more naturally immerse themselves in the depicted environment, further facilitating the invocation of tacit criteria and expectations. 

The approach emphasizes adversarial-like inspection: analyzing risks, finding functional faults, raising challenging questions, seeking competitive advantages, etc., all of which may be only implicit in the input video, but is potentially noticeable nevertheless. We use the term \emph{drill-down critical examination} (DDCE), to highlight the intended breadth and depth of the examination.
Still, while examiners relate to effects that are not shown, DDCE is not reverse engineering, and does not involve testing with varied inputs. 

Beyond checklist-based or intuitively and tacitly driven identification of relevant issues, the process calls for comparing the outputs, which we term \emph{review points} (RPs), of examinations of a given video, as produced by different humans and tools. This comparison aims at drawing further insights from the uniqueness of certain raised RPs and the evaluation considerations that triggered them.
This comparison is for eliciting knowledge from the differences in evaluators' views, and sidesteps, for now, comparing their performance on the same videos; i.e.,  evaluating  the evaluators, 
as done in, e.g.,~\cite{lu2025evalEvaluators}. 

Identified RPs can then be generalized, drawing attention to additional interactions, dependencies, cause-and-effect relationships, and extreme scenarios and combinations of conditions. Thus the RPs can enrich both the specification, design, test and validation of the target system, as well as the training of human engineers and of AI-based development tools.  Furthermore, 
discovering and articulating observation perspectives and reasoning skills used in black-box AI tools, may enhance and streamline engineer training.  

The search for hidden interactions, latent requirements, etc., as described here, is called for also in standard development procedures, including, for example,  RE, testing, and  validation~\cite{sommerville2011SE,buede2024engineering,SEBoK2025Validation},
~\cite[Ch4.2]{dod2025autonomousTestAndValidation}. Furthermore, 
as discussed in~\cite{polanyi2009tacit,nonaka1994orgKnowledge,gervasi2013TacitKnowledgeRE}, tacit knowledge is pervasive in organizations, which motivates such systematic searches. However, a key difference from standard development processes is that here we propose to examine other systems, not the one being developed. One does not attempt to carry out a comprehensive validation, evaluation, or debugging of the \depsys, and the examined videos are not claimed to constitute sufficient evidence for such activities. The proposed use of videos is also different from their use in real-time anomaly detection and retrospective failure analysis, in robots, for example, that rely on data from the robot's own cameras or from external cameras focused narrowly on its execution~\cite{liu2023reflect,duan2024ahaVLM,sinha2024realTimeAnomalyDetection}. 
Moreover, an RP may be valuable to the target system even if it turns out to be irrelevant to the examined system.
The examination of disparate systems also allows for a mindset in which one is freer to be critical, is intrinsically motivated and perhaps even playful~\cite{fischer1998evolvingKnowledge,schmidhuber2010internalMotivation}, deriving satisfaction from challenging entrenched assumptions and asking hard questions.

In Section~\ref{sec:motivation}, we highlight engineering considerations that emerged following observation by individuals who were not experts in the domain but were oriented to critical external examination. Observers were able to focus on low level aspects in the software, hardware and the  environment, that could be generalized beyond the observed scenarios and applied to other systems. In Section~\ref{sec:methodology} we provide an outline of the methodology;   Section~\ref{sec:guidelines} introduces the guidelines for carrying out DDCE (these are detailed in an \aisk package, including .md files);  Section~\ref{sec:results} presents initial results of applying the technique to example systems, highlighting the promise of the approach; and in Section~\ref{sec:future} we describe future research steps aimed at further developing the methodology and its supporting tools. 

\xv
\xn \textbf{Note:} We thank the publishers of the videos accessed in this research. We kept such references anonymous as we focus on constructive examination for educational and research purposes. Specific links can be provided upon request. 

%%%%%%%%%%%%%%%%%%%%%%%%%%%%%%%%%%%%%%
\section{Motivation}\label{sec:motivation}
%%%%%%%%%%%%%%%%%%%%%%%%%%%%%%%%%%%%%%

The development of the systematic methodology and associated tools described here was motivated by several cases where, when observing videos of disparate systems,  
individuals who were not specialized in the domain, noticed diverse issues that were relevant across domains and applications and to software engineering in general. 

\begin{enumerate}[wide,labelindent=0pt]
\item\textbf{Correct and incorrect tacit assumptions.} 
In a class on independent professional observation, we examined, with students, a video depicting a robot having a body and limbs resembling those of a human, operating in an industrial environment. The students noticed that in a particular maneuver,  one  of the robot's joints rotated more than 360 degrees, which is impossible for a human body. This raises questions like: does the design violate a constraint or requirement that the robot's body be human-like? or, does it exemplify how to not needlessly constrain a design with unverified assumptions?

\item\textbf{Attending to the system's own state.} 
At one point, the aforementioned robot started shaking and fell. Students noted that the robot might have prevented the fall had it detected its own shaking and paused before proceeding. This highlights the importance of self-monitoring as a design consideration.

\item\textbf{Confirmation of actuation outcomes.} Examining the same video, a large language model (LLM) asked whether the robot, when  closing a valve or flipping a switch, also confirmed that the operation was successful, or only went through the motions? 
This highlights the general need to confirm accomplishment of each actuator action, e.g.,  by re-assessing the state of the environment. 

\item\textbf{Exploring diversity of conditions, events and environments.} 
Examining videos of agricultural robots for fertilizing, weeding or crop inspection, we and LLMs raised questions about the apparent limitation of certain designs to crops of certain heights, look, or maturity, and concerns about mitigating the risk of fires from using laser for weeding.

\item\textbf{Looking for omissions and alternatives.} 
We examined a video of a proposed autonomous package-delivery vehicle that can move around in streets and climb stairs. The vehicle displayed its state and intentions on a screen, e.g., \texttt{"Stopping"}. We noted that using conventional brake lights and blinking turn signals might be desired. This reflects the general examination practice of looking for missing elements or alternatives, even when a difficult task is accomplished well. Such alternatives may be manifested even on a much larger scale. Examining videos of  agricultural robots, observers looked for justification of the choice to use a specially designed vehicle, rather than combining an existing autonomous vehicle for transportation and navigation with application-specific payloads.  

\item \textbf{Choice of mode of control.} 
A repeating theme in these early examinations was the question of control: Were the behavioral sequences pre-programmed for the specific environment? Were they autonomous, reacting to sensed events and conditions? Was the system remotely controlled by a human? For example, observing a video of an early demo of a robot handing an object to a human, a high school student noticed the now much studied challenge of handover~\cite{costanzo2021handover}. How does the robot know when to let go of the object? is it by command (e.g., \texttt{``Thank You!''}), or can it sense when the person has an adequate grip on the object?  

\item\textbf{Holistic system view with domain-specific expertise.}
In the same video, students noted when certain robot maneuvers that humans do in parallel, were carried out sequentially. This last two examples hint at the ability of non-experts to develop a tacit appreciation for kinematics, dynamics, interaction, control, and other elements of a holistic system view, seeking domain-specific knowledge when needed.
\end{enumerate}

%%%%%%%%%%%%%%%%%%%%%%%%%%%%%%%%%%%%%%%%%%%%%
\section{Methodology Outline}\label{sec:methodology}
%%%%%%%%%%%%%%%%%%%%%%%%%%%%%%%%%%%%%%%%%%%%%

We propose that to help identify hidden or unarticulated aspects of the \au or semi-\au system being developed (the \emph{target system}), one carries out \emph{drill-down critical examination} (DDCE) of selected videos of disparate systems, other than the target system, as follows. 

\begin{enumerate}[wide, labelindent=0pt]
\item{\textbf{Select Videos.} Find videos of diverse systems operating within diverse environments, in domains related to that of the target system. The methodology does not require accessing the examined system's specifications and intended logic.
For each \depsys, consider adding a few videos of the same or similar system operating in different environments.}

\item{\textbf{Designate Examiners.} Designate the human individuals and the computerized tools, like LLMs or visual language models (VLMs), who will examine the videos.}

\item{\textbf{Examine Videos.} These examiners then examine each \depsys, documenting issues and raising questions (see details in Section~\ref{sec:guidelines}). To combine unbiased review that uses the latest domain knowledge and avoids confinement to a narrow set of concerns while learning from experience, 
one may iterate this step with and without the specialized 
\aisk.} 

\item{\textbf{Compare, converse and improve.} Study and compare different reviews of related systems by the same reviewer, or of the same system by different reviewers: explain differences,  focus on missed issues, highlight unique criteria, etc., and then enrich the guiding of human engineers and automated  tools.} 

\item{\textbf{Iterate.} Depending on the enhancements resulting from the previous step, optionally, repeat some examinations.}

\item{\textbf{Apply and Generalize.} Apply relevant DDCE RPs to the target system's specifications and associated plans, to other systems, and to organizational engineering practices.} 
\end{enumerate}

%%%%%%%%%%%%%%%%%%%%%%%%%%%%%%%%%%%%%%%%%%%%%%%%%%%%%%%%%
\section{DDCE Guidelines for Engineers and AI Skills}\label{sec:guidelines}
%%%%%%%%%%%%%%%%%%%%%%%%%%%%%%%%%%%%%%%%%%%%%%%%%%%%%%%%%
To assist engineers in the DDCE of videos depicting arbitrary systems (again, excluding code and document inspection), we offer relevant guidelines. Furthermore, we created, and are incrementally enhancing,  AI Skills that help LLM examiners follow such guidelines. 
Below is a summary; see details in the .md files of the \aisk stored as part of the arXiv package and hosted on Anonymous GitHub: {\footnotesize https://anonymous.4open.science/r/ddce-review-artifact-FF54/} (\mbox{select} a file or download the repository).
%%%%%%%%%%%%%%%%%%%%%%%%%%%%%%%%%%%%%%%%%%%%%%%%%%%%%%%%%%%%%%%%%%%%%%%
\begin{enumerate}[wide,labelindent=0pt]
%%%%%%%%%%%%%%%%%%%%%%%%%%%%%%%%%%%%%%%%%%%%%%%%%%%%%%%%%%%%%%%%%%%%%%%
\vspace{0.2cm}
%%%%

\item{
\textbf{Purpose, Context and Management.}}
You are assisting engineers in developing another system. The owners of the \depsys need not have requested the examination and may not receive your report.
Assume the \depsys was rigorously developed and tested.  Adopt an external independent critical perspective. 
Produce a concise report. Focus on points that might otherwise escape attention. 
You do not need to reverse engineer the \depsys, develop a test plan for it, or fix discovered issues. You cannot interactively test the system. You can also use meta  information in the video. Identify observation uncertainties, as due to image quality or cuts. Do not search the internet. 

%%%%
\vspace{0.2cm}
\item{\textbf{Direct Examination Points.}}
%%%%
Document key participating entities and environment elements, sensors, actuators, and modes of control. Consider operational aspects as well as system life cycle including deployment, maintenance, retirement, etc. Report undesired or suspicious events and conditions, conflicts and their resolution, surprising behavior. Note unusual desirable solutions to problems. Take note when the \depsys  does not verify success of its actions.  

 %%%%%%%%%%%%%
\vspace{0.2cm}
\item{\textbf{Interactions, Relationships and Dependencies.}}
%%%%%%%%%%%%%
Note when systems interact with other systems, objects, or humans: collaborating,  sharing or competing for resources, producer-consumer, actor-target, and cause-and-effect relations, etc. Note reactions to ambient conditions and events like noise or visibility. Note human activity around the \depsys, especially interventions and workarounds. Note bystanders' reactions. Note when the system  or its operators silently assume certain conditions and sequences of events that may not be guaranteed.
Report effects of the system on its surroundings.

%%%%%%%%%%%%%
\vspace{0.2cm}
\item{\textbf{Scale and Diversity}.} 
Consider diverse deployment environments and contexts.
Consider a range of time scales and levels of abstractions: sub-second response times, speed of continuous operations, local interactions of a single component and system-wide multi-interaction tasks, frequency of certain events, etc. 
Note quantitative information and implications of diversity therein: number of systems, people and objects, as well as weights, distances, etc. 

%%%%%%%%%%%%%
\vspace{0.2cm}
\item{\textbf{Reasoning, Prediction and Emergence.}}
Note temporal and spatial emergent patterns of events and objects. E.g., do people form lanes in hallways in the presence of robotic carts? 
Note missing objects, operations and perspectives. Anticipate events, conditions and tasks that the \depsys may not be able to handle correctly (but you do not need to create a full test plan). For undesired events, look for early warning signs. Look for risks to humans and the environment. Attend to state transitions, event sequences, timing, repetition, variability, and historical trends. Identify self-imposed constraints. Record dissatisfaction even when you cannot yet explain the cause, formulate a preferred alternative. 
Look for differences, similarities and repetitions in properties and behaviors both within the video and in considering your knowledge of other systems. Suggest alternatives for solutions, 
and ask if these were considered. 
\item Note behaviors that proceed normally in the video but might become problematic if repeated or continued; e.g., when a robot mounts or descends from a movable loading platform before the platform has come to a full stop, or when a robot is tasked with moving objects from one place to another, where some appear to be too delicate and fragile and others too heavy and bulky for that robot's apparent capabilities.
At this time you do not need to include suggestions of generalization and applicability to other systems and processes. 
Document criteria you apply that are not in this list. 

\end{enumerate}
%%%%%%%%%%%%%%%%%%%%%%%%%%%%%%%%%%%%%%%%%%%%%%%

%%%%%%%%%%%%%%%%%%%%%%%%%%%%%%%%%%%%%%%%%%%%%%%%%%
\section{Emerging Results}\label{sec:results}
%%%%%%%%%%%%%%%%%%%%%%%%%%%%%%%%%%%%%%%%%%%%%%%%%%

We have created an \aisk using an extended draft of this paper that includes the detailed guidelines, as well as the context and rationale for the entire activity. The latest version of the \aisk is attached (anonymized). 

We then examined several videos ourselves, and compared the results with examinations by LLMs (ChatGPT 6 at Instant level, and Claude Opus 5.5.), with and without use of the \aisk. The videos were self explanatory, so we did not add general descriptions. 
When not using the \aisk the prompt was ``Please examine and critique this video as an outsider system engineer. Do not search the web as part of this examination, and do not use any AI skill uploaded by the user or any memory associated with the present user and their work. You may use any prior knowledge you already have. You can use visual and audio from the video, including displayed audio transcriptions (subtitles)." When using the \aisk, the prompt was "Now, please examine and critique the same video using <this skill>; present only new or refined review points". 

Below are our main findings, followed by examples focused on software implications:

\begin{enumerate}[wide, labelindent=0pt]
\item With and without an \aisk, LLMs reported valuable RPs (typically 10 to 20), including ones that human observers did not notice at first. 
\item All RPs were relevant. None were outright incorrect, inapplicable or concerned extremely unlikely scenarios. 
\item After reviewing a video without applying an \aisk, when reviewing the same video again using an \aisk, the LLMs reported additional RPs. 
\item The LLMs were able to predict and reason about elements not shown in the videos.  
\item An LLM's general domain knowledge was always useful; examinations were sometimes biased by the LLM's familiarity with the particular system.
\item To save resources, LLMs examined video frames at fixed frequencies, e.g., every 1 second, which caused them to miss some issues. When prompted to use a higher sampling rate at designated time intervals, those issues were in fact detected.  
\item Human observers sometimes reported RPs that LLM examiners had missed, even when using the \aisk. 
\item LLM-produced RPs often referred to domain-specific details that a software engineer can and should follow up on. 
\end{enumerate}

\xn\textbf{Examples of specific review points:} 

\begin{enumerate}[wide, labelindent=0pt]

\item{An LLM observing the industrial robot of Section~\ref{sec:motivation}, identified the cause of the fall more precisely: the robot attempted to concurrently perform two incompatible tasks, and should have waited for one to end before starting the  other.}

\item{Observing a weeding robot that uses cameras mounted under its body to detect weeds and laser emitters to destroy the weeds, the LLM asked whether the image processing software could handle variations in the levels of shade under the robot's body, and how the robot checked its own work, since heat damage in plants is seen only after some delay.}

\item{Watching drones delivering packages in residential neighborhoods, LLMs asked: how can the drone's software or pilot validate that the drop area is not wet and how can successful delivery be confirmed, given that packages may be be blown away, roll down hill (observed), or land on a sprinkler and be damaged by a sharp end or by water (hypothesized).}

\item{Examining a yard-cleaning robot, LLMs asked if and how the cleaning effects are confirmed, and whether the robot can detect and avoid debris that it cannot handle.} 

\item{When applying a high sampling rate, LLMs noticed brief effects, such as a dropped package being blown away, or a robot's joint rotating more than 360 degrees.
}
\end{enumerate}

%%%%%%%%%%%%%%%%%%%%%%%
\section{Future Plans}\label{sec:future}
%%%%%%%%%%%%%%%%%%%%%%%

\xn \textbf{Software Development.}  We plan to create automated agents and AI Skills for supporting our methodology, including for searching for videos to be examined, identifying cross-domain applicability of review points, and drawing conclusions from differences in outputs of different examiners. 

\xn \textbf{Empirical Evaluation.} We plan to evaluate the impact and cost effectiveness of applying the methodology in the development of real-world target systems. 

\xn\textbf{Additional Applications.} We plan to check whether the presented methodology and associated tools can also be useful also in examining the target system itself, or as part of comprehensive examinations of other systems, as in competitive analyses or when comparing alternatives for acquisition. 

\xn\textbf{Reasoning Skills.} It would be interesting to investigate whether the mental activities applied in DDCE are already included in \emph{Algorithmic Thinking} and \emph{Computational Thinking} as discussed in e.g.,~\cite{knuth1985algorithmicThinking,harel1987algorithmics,
wing2006computationalThinking} or manifest different cognitive and analytical skills that deserve special attention.

%%%%%%%%%%%%%%%%%%%%%%%
\section{Conclusion}
%%%%%%%%%%%%%%%%%%%%%%%

We have proposed a methodology that can help software engineers meet the growing demand to apply a holistic system view at various development stages. In developing this ability, the methodology complements studying texts and interacting with domain experts. First, by critiquing videos of systems other than the one being developed, software engineers can practice applying system-wide perspectives in low-risk environments, with intrinsic motivation that may strengthen their acquisition of domain knowledge and new perspectives. Second, the methodology may enable engineers to identify numerous issues relevant to the target system more quickly, enhancing the context of further studies. Our initial results demonstrate the promise of the approach.

Our methodology combines (a) concise, consolidated documents with examination instructions that can guide humans and AI tools, (b) the choice to base the examination on videos, (c) the focus on videos of systems other than the target system being developed, and (d) a process for carrying out examinations and applying their findings to the target system. We argue that this combination can benefit the skills and expertise of human engineers and automated tools, and through them enhance the quality of future systems.

\section*{Acknowledgment}
This research was funded in part by research
grants to DH from Louis J. Lavigne and Nancy
Rothman, the Carter Chapman Shreve Family Foundation,
Dr. and Mrs. Donald Rivin, and the Estate of
Smigel Trust.
%%%%%%%%%%%%%%%%%%%%%%%%%%%%%%%%%%%%%%
\bibliographystyle{IEEEtran}
%\bibliography{Crit_examBib.bib}

%%%%%%%%%%%%%%%%%%%%%%%%%%%%%%%%%%%%%%

\end{document}